\documentclass[sigconf]{acmart}

\usepackage{multirow}

\usepackage{float}

\usepackage{soul}      
\usepackage{xcolor}    
\sethlcolor{yellow}    

\usepackage{natbib}   

\soulregister\cite7   

\PassOptionsToPackage{protrusion=true,expansion=true}{microtype}
\usepackage{microtype}
\usepackage{xurl} 
\AtBeginDocument{%
  }

\copyrightyear{2026}
\acmYear{2026}
\setcopyright{cc}
\setcctype{by}
\acmConference[CHI EA '26]{Extended Abstracts of the 2026 CHI Conference on Human Factors in Computing Systems}{April 13--17, 2026}{Barcelona, Spain}
\acmBooktitle{Extended Abstracts of the 2026 CHI Conference on Human Factors in Computing Systems (CHI EA '26), April 13--17, 2026, Barcelona, Spain}
\acmDOI{10.1145/3772363.3798598}
\acmISBN{979-8-4007-2281-3/2026/04}

\begin{document}

\title[Phatic Communication and Algorithmic Contagion in Meme Sharing]{"Don't Mess Up My Algorithm":
Phatic Communication and Algorithmic Contagion in Meme Sharing}

\author{Ji Eun Song}
\orcid{1234-5678-9012}

\affiliation{%
  \institution{Seoul National University}
  \city{Seoul}
  \country{Republic of Korea}
}
\email{jieun.song@snu.ac.kr}

\author{Hyunsoo Jang}
\affiliation{%
  \institution{Seoul National University}
  \city{Seoul}
  \country{Republic of Korea}
  }
\email{gnfk0119@snu.ac.kr}

\author{Juhee Im}
\affiliation{%
  \institution{Seoul National University}
  \city{Seoul}
  \country{Republic of Korea}
  }
\email{jhdsny1105@snu.ac.kr}

\author{Joongseek Lee}
\affiliation{%
  \institution{Seoul National University}
  \city{Seoul}
  \country{Republic of Korea}
  }
\email{joonlee8@snu.ac.kr}

\renewcommand{\shortauthors}{Song et al.}

\begin{abstract}
On algorithmic social platforms, exchanging memes via direct messages (DMs) serves as phatic communication that affirms relationships, yet users often interpret these exchanges as signals shaping personalized recommendations, creating tension between relational practice and algorithmic control. This study examines how users perceive DM meme exchanges on Instagram rather than auditing Instagram's underlying recommender mechanisms, and how beliefs about DM-recommendation linkages shape coping strategies and feelings of powerlessness. We conducted semi-structured interviews with 21 active meme-DM users. Participants classified memes as recipient-friendly or recipient-unfriendly based on relational fit; many described the spread of unfriendly memes as “algorithmic contagion.” Controls were constrained by relational norms, low perceived efficacy of feedback tools, and opaque DM-recommendation linkages. We articulate how DM-based relational practices entangled with personalization infrastructures and propose three design implications: transparent linkage explanations, conversation-level opt-outs, and conservative learning that down-weights DM-originated signals.
\end{abstract}




\maketitle

\section{INTRODUCTION}
Meme exchange on social media is often less about information than phatic communication—acknowledging presence and aligning affect. With a short caption and image, memes can signal “you find this funny” or “you noticed what I’m going through,” offering more texture than a like or emoji while remaining lighter than extended conversation, thus supporting everyday relational maintenance.

This practice is especially active in the DMs of algorithmic platforms, where short-form content is increasingly shared and interpreted in private. DMs function as a backchannel that runs alongside the algorithm-curated public feed. Because many platforms use hybrid recommenders combining preference-based and social signals, users may believe DM meme exchanges shape recommendations—creating tension between phatic relational goals and the desire to “not mess up my algorithm.”

We examine how users make sense of DM meme exchange as phatic communication within perceived hybrid recommender environments, and how these beliefs shape algorithmic control and coping. We do not infer platform mechanisms; we analyze users' folk theories about DM–recommendation linkages. We conducted semi-structured interviews with 21 young adults in Korea who actively exchange memes via Instagram DMs. We contribute a recipient-centered typology, an account of DM–recommendation folk theories, and design directions for relationship-preserving control.
\section{RELATED WORK}

\subsection{Meme as Phatic Communication}
Malinowski \cite{Malinowski1943} defines phatic communication as talk where maintaining social ties matters more than conveying information. Memes—remixed, iterative messages that circulate rapidly in participatory culture \cite{Wiggins2014Bowers}—often serve this phatic role. Like other lightweight visual tokens (e.g., GIFs, emojis) \cite{Peng2025Chen}, memes can signal intimacy and relational stance, sustaining interaction even when their literal content is minimal. While early meme scholarship often treated memes as sites of public discourse and cultural contestation  \cite{shifman2013memes, milner2018world, knobel2007online}, research on everyday exchange highlights memes as vehicles for connected presence: low-effort communication producing belonging, affective ties, and social warmth. Meme sharing can also transmit "insider codes" and perform shared identity \cite{newton2022more}, echoing Gal’s \cite{gal2019ironic} point that ironic online humor draws social boundaries and functions as social glue through exclusivity and mutual recognition. Recent work foregrounds memes’ phatic functions in private channels (DMs, group chats) \cite{brody2023meme}, where memes help users acknowledge co-presence, reinforce intimacy, and share affect. In these contexts, responses are highly sensitive to alignment with the recipient’s tastes and identity \cite{green2016social}, so memes can create either relational safety or subtle discomfort. Across qualitative and quantitative studies \cite{shamayleh2025digital, shandilya2022need, ali2023getting}, meme exchange is most frequent within strong ties (friends, partners, family), reinforcing that memes operate as relational—not informational—interaction.

\subsection{Direct Messaging as a Backchannel}
In-app DMs can be understood as a digital backchannel that runs alongside the “frontchannel” of algorithmically curated feeds. Building on HCI work on backchannels in conferences and classrooms \cite{mccarthy2005digital, toledo2010educators} and private moderation backchannels in online communities \cite{seering2024chillbot}, we extend this lens to DM-based meme exchange that accompanies—and can subtly reshape—feed experiences. On platforms such as Instagram and TikTok, users often share memes or news links with close friends via DMs rather than public reposts, shifting both diffusion and interpretation into private threads where people add context and discuss feed content \cite{lottridge2018let, sujon2017public}. DMs also support sensitive disclosure and support-seeking among teens \cite{ali2022understanding, huh2023help}, and enable coordinated algorithmic action: K-pop fan groups organize liking, sharing, and replay campaigns through group chats (e.g., WhatsApp, Twitter DMs) to boost idols' rankings \cite{xiao2025let}.

\subsection{Algorithmic Awareness and Coping Experience: Understanding, Control, and Helplessness}
A large body of work examines how users understand and respond to algorithmic recommenders on social and content platforms. Studies of Facebook News Feed, YouTube, and TikTok show many users do not fully grasp algorithmic curation: some assume they “see everything,” while others rely on folk theories that conflate watch history, popularity, and advertising policies \cite{eslami2015always, eslami2016first, rader2015understanding, alvarado2020middle}. Beliefs range from trust that the algorithm “knows me” to suspicion that it filters certain identities an “identity strainer”; minoritized users may perceive suppression and use hashtag tuning or mutual “boosting” \cite{karizat2021algorithmic}. DeVito conceptualizes this sensemaking as adaptive folk theorization \cite{devito2021adaptive, devito2017algorithms}. Bucher shows how algorithmic imaginaries and emotions co-produce everyday practice and collective resistance (e.g., \#RIPTwitter) \cite{bucher2019algorithmic}. Yet control attempts (e.g., reporting, blocking, “Not Interested”) often feel opaque and ineffective, producing fatigue, demands for autonomy and transparency \cite{mozilla, hong2025social, oeldorf2023exploring}, and digital resignation \cite{draper2019corporate}; only some users develop coping tactics through algorithmic literacy and “algorithmic gossip” \cite{bishop2019managing}. Transparency interventions mostly target public feeds: even brief explanations can shift perceived legitimacy and accountability \cite{rader2018explanations}. In contrast, users strongly resist personalization based on private data (e.g., DMs) \cite{kozyreva2021public}, and highly personalized systems like TikTok can make small interactions feel like feed “contagion,” triggering cascades of unwanted similar content and loss of control \cite{vera2025they}.

Prior work has described how highly responsive recommenders can make small interactions feel like feed ‘contagion.’ We extend this line by locating the trigger in phatic exchanges in private DMs, showing how relational obligations reshape perceptions of algorithmic risk and constrain control.
\section{METHOD}
\subsection{Participant Recruitment and Sampling}
We recruited adults (19+) based in Korea who had exchanged at least three memes via DM in the past week. Recruitment occurred through university communities, social media posts, and referral-based snowball sampling. We define memes as shareable, humor-, irony-, or affect-oriented visual media exchanged via DMs—specifically captioned image macros, screenshots, and short-form video memes that draw on recognizable templates, trends, or remixing practices, and that participants themselves regard as “memes.” In the screening survey, we first provided an operational definition of “meme,” then asked about participants’ DM meme-sharing practices and whether they believed DM meme exchanges affected their algorithmic feeds. These questions prompted reflection and helped participants articulate concrete experiences. We interviewed 21 participants, prioritizing those who frequently exchanged memes and those likely to be informative for exploring emerging theoretical concepts. (see Table 1 for a summary). Participants received USD \$25 compensation.

\begin{table*}[t]
    \centering
    \caption{Demographics of interview participants}
    \small
    \label{tab:table1}
    \begin{tabular}[\textwidth]{p{0.8cm}p{0.8cm}p{1.3cm}p{3.0cm}p{0.8cm}p{0.8cm}p{1.3cm}p{3.0cm}}
\toprule
\textbf{ID} & \textbf{Age} & \textbf{Gender} & \textbf{Occupation} & \textbf{ID} & \textbf{Age} & \textbf{Gender} & \textbf{Occupation} \\
\midrule
P01 & 30 & Male & Advertising Planner & P12 & 32 & Female & Environment Activist \\
P02 & 28 & Female  & Graphic Designer & P13 & 27 & Female  & Graduate Student \\
P03 & 26 & Female  & Graduate Student & P14 & 22 & Male  & Undergraduate Student \\
P04 & 28 & Male & Start-up Founder & P15 & 28 & Male  & Patissier \\
P05 & 23 & Male  & Undergraduate Student & P16 & 24 & Female  & Undergraduate Student \\
P06 & 20 & Male  & Undergraduate Student & P17 & 24 & Female  & Undergraduate Student \\
P07 & 27 & Male  & Graduate Student & P18 & 27 & Female & Advertising Salesman \\
P08 & 22 & Female  & Undergraduate Student & P19 & 28 & Female  & Academic Staff \\
P09 & 25 & Female  & Undergraduate Student & P20 & 25 & Female  & Undergraduate Student \\
P10 & 26 & Male  & Graduate Student & P21 & 29 & Female  & STEM Researcher \\
P11 & 24 & Female & Undergraduate Student \\
\bottomrule
\end{tabular}

\end{table*}

\subsection{Data Collection and Analysis}
We conducted semi-structured interviews via Zoom, each lasting 60–90 minutes. The interview had two parts: (1) participants’ general experiences of meme sharing and (2) experiences of perceived algorithmic influence and control following phatic communication. We did not collect third-party identifying information; instead, we encouraged participants to review memes they had recently exchanged and use them as prompts when answering questions. The interview protocol was iteratively revised as concepts emerged during analysis. All interviews were recorded and transcribed.

We followed a grounded theory approach \cite{charmaz2012qualitative, corbin2014basics}, collecting and analyzing data in parallel. After interviewing 21 participants (N=21, F=13, mean age=25.95), we determined that data saturation had been reached. Using analytic memos written throughout the process, we conducted line-by-line thematic coding after each interview. Coding proceeded bottom-up: we began with open coding, then refined categories around (a) phatic types of received memes and (b) algorithm-related experiences grounded in participants' folk theories. We paid particular attention to conflicts between phatic relational goals and attempts to control algorithmic outcomes. To ensure analytic rigor, all authors participated in coding and engaged in iterative discussion and consensus-building.

\section{FINDINGS}

\subsection{Meme DMs as Phatic Communication: Recipient-Friendly vs. Recipient-Unfriendly}

Across participants’ accounts, exchanging memes in DMs was understood as a relationship-centered phatic practice. Rather than a means of sharing new information, meme DMs were interpreted as lightweight ways to check in, signal presence, and reaffirm intimacy. Participants differentiated these exchanges based on how well a meme “fit” the recipient, yielding two broad types: recipient-friendly and recipient-unfriendly memes.

Recipient-friendly memes appeared in three forms: (1) context-anchored, referencing specific topics or narratives derived from real-life interactions —“It was this thing we’d talked about when we met up in person, and then I got it as a short video—like a Shorts edit. So I was like, ‘Ohhh, that’s the part we were talking about.’” (P08); (2) identity and taste alignment, matching the recipient's preferences—“I think my friend sent it because they thought I'd like it. And yeah—I did.” (P09); and (3) consensual safe banter, functioning as low-stakes joking within mutually understood boundaries—“Just to tease me—like, to mess with me.” (P13). In contrast, recipient-unfriendly memes included: (1) context-missing memes that felt ill-timed or irrelevant—“There's literally no context. They just keep sending this same photo. I don’t even know why.” (P18); (2) taste-misaligned memes experienced as burdensome or uncomfortable due to a preference mismatch— “Yeah, I’ve gotten memes I’m just not into.” (P04); and (3) harmful/boundary-violating memes containing objectionable content or intruding on psychological or social boundaries—“My friend kept sending scary stuff, and it made me uncomfortable. And honestly, it kind of scared me.” (P18)

\subsection{Perceived Influence of Received Memes on Recommendation: “Expanded Exploration” vs. “Algorithmic Contagion”}
Many participants believed that content shared via DMs could influence subsequent recommendations. Our focus is not on the technical mechanism of this linkage, but on how participants perceived and made sense of it. Participants often noticed similar content reappear in their feeds after receiving a meme via DM. They believed that additional interactions—opening the link, watching a video to the end, saving it, or liking it- were interpreted by the algorithm as preference signals, expanding related recommendations or biasing the feed.

These perceptions varied according to the recipient-friendliness of memes. When participants received friendly memes, they perceived these interactions as leading to more aligned recommendations—a positive framing we term \textbf{expanded exploration}. “I want my algorithm to show me more of what I actually like. Lately I got into cow and goat stuff and I was hoping it’d show up more. Goats weren’t really popping up, but then my friend started sending me goat content, and after that it showed up way more. I liked that—it was stuff I wanted to see.” (P17) In contrast, when participants received unfriendly memes, they believed that such content propagated into their feeds, describing the resulting feed as “spilled over”, a negative framing we term \textbf{algorithmic contagion}. “It felt like my friend was doing it with bad intentions, so I just blocked more stuff. My friend was like, ‘I can’t be the only one who feels gross about this,’ and it felt like they were trying to pass it on to me... so I purposely hit ‘Not Interested’ even more.” (P13)

\subsection{Feeling Powerless to Intervene in a Spilled-over Feed}
Low-cost feedback (e.g., occasional ``Not Interested'') felt ineffective, while higher-cost actions (e.g., blocking or leaving threads) carried relational penalties; as a result, most participants tolerated unwanted content. Participants often reasoned about senders’ intentions (e.g., care, teasing, or ‘passing on’ uncomfortable content), suggesting that spillover concerns are socially constrained within dyads rather than managed individually (e.g., withholding reactions, changing topics, or jokingly deflecting). Although participants evaluated the spread of unfriendly memes negatively, their interventions were limited. A few attempted to separate accounts to maintain a “clean” feed, but the ongoing effort required to manage multiple accounts made this an exceptional strategy. Most participants (N=18) used "Not Interested" only on specific posts or simply scrolled past.

This low level of intervention reflected three-layered constraints. At the system level, participants could not tell whether “Not Interested” labels or reports meaningfully affected recommendations, weakening perceived efficacy. I blocked and blocked, and I kept doing the ‘I don’t want to see this post’ / ‘Not Interested’ thing too—mostly for idol posts, I think. I even went out of my way to search for those meme accounts and block them. But even then, if I let my guard down and watch just a few, some new meme account starts invading my algorithm again. That’s when I was like, ‘Okay, none of this works—I need a different approach.’” (P12) At the normative level, leaving a DM thread or blocking someone was seen as damaging to relationships and at odds with the phatic function of DMs. “I mean, they’re still my friend—how am I supposed to block them? I just put up with it.” (P13) Finally, at the content level, memes were treated as ephemeral and low-stakes, making granular management feel not worth the effort—“the hassle outweighs the benefit.” “It wasn’t, like, hateful or super gross—nothing that shouldn’t show up. If it were that bad, I’d probably hit ‘Not Interested,’ but it wasn’t quite at that level.” (P17) As a result, participants remained in a resigned state, continuing to encounter unwanted memes while feeling that “this is something I just have to put up with.”

\section{DISCUSSION}
Extending prior accounts of memes as “social glue” \cite{shifman2013memes, milner2018world, gal2019ironic} to DM backchannel, we show that DM meme exchange operates as recipient-centered phatic tie maintenance: recipients judge affect and identity fit, distinguishing friendly from unfriendly memes, so the same content can feel like relational safety or a burden. Echoing the identity strainer \cite{karizat2021algorithmic}, participants framed friendly DM memes as expanded exploration, whereas unfriendly memes were experienced as algorithmic contagion—content that spills into feeds and prompts avoidance. Participants' algorithmic imaginaries cast the DM backchannel as probabilistic belief throuogh algorithmic abstraction, echoing Parisi' \cite{parisi2013contagious}s notion of contagious architecture. In this space, contagion is not a physical fact, but an abstract certainty imposed on the user by incomputable data.  Yet participants rarely took strong action beyond occasional “Not Interested” (or, rarely, account separation), citing low perceived efficacy of feedback tools \cite{hong2025social}, relational costs of blocking or leaving chats, and memes' low-stakes ephemerality—patterns consistent with digital resignation \cite{draper2019corporate}. This tension—phatic exchange strengthening ties yet imposing perceived costs on recommendations—motivates three design directions: (1) relationship-friendly controls decoupling phatic interaction from preference learning, (2) explanation interfaces making DM–recommendation linkages visible and actionable, and (3) conservative learning mitigating harm even under default non-intervention.

\subsection{Relationship-Friendly Controls that Preserve Phatic Communication}
Platforms should decouple relational maintenance from algorithmic learning. Concrete controls might include: (1) a per-conversation or per-sender toggle such as "Do not use interactions in this chat for recommendations," and (2) a relationship-preserving option such as “Maintain this connection, but reduce recommendations derived from content shared by this person.” These patterns would allow users to continue phatic meme exchange while selectively reducing the feed effects of unfriendly memes.

Tradeoff: conversation-level opt-outs may reduce friend-based discovery and overall personalization utility, and may interact with safety/abuse detection pipelines that rely on private interaction signals; careful defaults and safeguards are needed.

\subsection{Expanding Explainability and User Negotiation for DM–Recommendation Linkages}
Participants reacted strongly to DM-driven spillover yet lacked visibility into how DM interactions shaped recommendations. This highlights a design need for socially safe controls: mechanisms that reduce algorithmic impact without signaling distrust or rejection to the sender by default. Platforms could pair contextual explanations with controls \cite{hong2025social, rader2018explanations}. For example, a feed card could disclose: “Recommended because it is similar to content you engaged with via private shares,” alongside an action such as “Reduce recommendations from this DM thread” (or “Reduce DM-based recommendations”). Given users’ discomfort with private-data personalization \cite{kozyreva2021public}, disclosures should favor privacy-preserving phrasing (e.g., not naming the sender unless users opt in) while remaining actionable.

Tradeoff: making DM-based linkages legible can inadvertently reveal sensitive relational information; explanations should be privacy-preserving, minimally specific by default, and allow opt-in granularity.

\subsection{Conservative Learning and Contagion Mitigation Under Default Non-Intervention}
Platforms should prioritize conservative learning that reduces harm even when users do not intervene. For instance, platforms could assign lower default weights to signals from phatic channels like DMs, promoting them only when repeated engagement with the same sender or content type is detected. This would limit rapid "spill over" from a few unfriendly memes and alleviate the algorithmic powerlessness observed in DM-based exchange.

Tradeoff: conservative weighting may slow beneficial learning for users who intentionally use private sharing for discovery; adaptive thresholds and user-facing tuning may balance caution with responsiveness.
\section{LIMITATIONS AND FUTURE WORK}
Limitations include a Korea-based young-adult sample (N=21; university/snowball recruitment) and an Instagram-only focus, which may limit generalizability. Future work should test other ages/cultures/platforms (e.g., TikTok), examine tie-strength and sender/collective dynamics, and validate DM–recommendation linkages with platform logs; here we report folk theories that may differ from system behavior.
\section{CONCLUSION}
This study reframes DM meme exchange as an intersection of phatic communication and perceived algorithmic signaling. Interviews with 21 participants show meme DMs function as relationship-centered check-ins, yet recipients distinguish friendly from unfriendly memes and widely believe DM-shared memes shape recommendations—describing misaligned spread as "algorithmic contagion." We outline relationship-preserving controls and conservative learning.

\bibliographystyle{ACM-Reference-Format}
\bibliography{references/bibliography}

\appendix

\end{document}